\documentclass[prl,aps,twocolumn,showpacs,nobibnotes,epsf]{revtex4}
\usepackage{graphicx}% Include figure files
\usepackage{dcolumn}% Align table columns on decimal point
\usepackage{bm}% bold math

\begin{document}
\draft

\title{Magnetic and Transport Properties in $CoSr_2Y_{1-x}Ca_xCu_2O_7$ ($x$=0$\sim$0.4)}
\author{X. G. Luo}
\author {X. H. Chen}
 \altaffiliation{Corresponding author}
 \email{chenxh@ustc.edu.cn}
\author{X. Liu, R. T. Wang, Y. M. Xiong, C. H. Wang, and G. Y. Wang}
\affiliation{Structure Research Laboratory and Department of Physics,\\
University of Science and Technology of China, Hefei, Anhui,
230026, P. R. China}
\author{X. G. Qiu}
\affiliation{State Key Laboratory of Superconductivity, Institute of Physics, Chinese Academy of Sciences,
Beijing 10080,  People's Republic of China}

\begin{abstract}
Magnetic and transport properties of
$CoSr_2Y_{1-x}Ca_xCu_2O_7$($x$=0$\sim$0.4) system have been
investigated. A broad maximum in M(T) curve, indicative of
low-dimensional antiferromagnetic ordering originated from
$CoO_{1+\delta}$ layers, is observed in Ca-free sample. With
increasing Ca doping level up to 0.2, the M(T) curve remains
almost unchanged, while resistivity is reduced by three orders.
Higher Ca doping level leads to a drastic change of magnetic
properties. In comparison with the samples with $x=0.0 \sim 0.2$,
the temperature corresponding to the maximum of M(T) is much
lowered for the sample $x$=0.3. The sample $x$=0.4 shows a small
kink instead of a broad maximum and a weak ferromagnetic feature.
The electrical transport behavior is found to be closely related
to magnetic properties for the sample $x$=0.2, 0.25, 0.3, 0.4. It
suggests that $CoO_{1+\delta}$ layers are involved in charge
transport
 in addition to conducting $CuO_2$ planes to
interpret the correlation between magnetism and charge transport.
X-ray photoelectron spectroscopy studies give an additional
evidence of the the transfer of the holes into the
$CoO_{1+\delta}$ charge reservoir.
\end{abstract}
\vskip 15 pt

\maketitle

\section{INTRODUCTION}
Charge transport and high temperature superconductivity(HTSC) is
believed to reside in the $CuO_2$ planes of all known HTSC
cuprates\cite{1}, except that $CuO_{1+\delta}$ chains have been
reported to participate in the b-axis transport of
$YBa_2Cu_3O_{7-\delta}$\cite{2}. In $YBa_2Cu_3O_{7-\delta}$ (also
denoted as $CuYBa_2Cu_2O_{7-\delta}$, Cu1212) there are two
different Cu sites, namely Cu(2) and Cu(1). Cu(2) resides in
superconducting $CuO_2$ planes and Cu(1) in $CuO_{1+\delta}$
chains. Any breach of integral $CuO_2$ stacks, even at macroscopic
level, affects superconductivity drastically. The $CuO_{1+\delta}$
chain acts as charge reservoir and provides the mobile carriers to
superconducting $CuO_2$ planes.

Some attempts have been performed successfully for complete
replacement of $CuO_{1+\delta}$ by other metal elements.
$MSr_2YCu_2O_{7+\delta}$(M-1212, M=Pb, Ru, Nb, Al, Hg, Fe, Ga, Ta
etc.)\cite{Shinesun,Roth, Isobe,Bauernfeind,ADACHI,Shimoyama} have
attracted extensive attention in the past. Some of them have been
made superconducting, such as Nb1212, Hg1212, Ru1212, Tl1212,
Ga1212 and Fe1212 \cite{Bauernfeind,ADACHI,Shimoyama}, and some
other, like Ta1212, remains nonsuperconducting. Another example,
Co1212, kept nonsuperconducting since it was synthesized firstly
in 1989\cite{Shinesun}, until very recently, Morita {\sl et al.}
made it superconducting(SC) by annealing Ca-doped Co1212 in
ultra-high $O_2$ pressure(5Gpa)\cite{Morita}. However, the reason
why the system is not SC before ultra-high pressure treatment and
it becomes SC after ultra-high pressure treatment, still
remains unknown.

Up to now, the research for Co1212 has been focused on its
structure and magnetic properties\cite{Huang, Awana, Awana1}. The
structure of Co1212 is derived from
$CuBa_2YCu_2O_{7-\delta}$(Cu1212), by completely substituting
divalent Cu(1) in $CuO_{1+\delta}$ charge reservoir of Cu1212 with
trivalent Co ion. In Cu1212 the charge-reservoir Cu atom with
surrounding oxygen atoms forms a square $CuO_4$ polyhedron.
However, the $CoO_4$ tetrahedral coordination polyhedron has been
reported in Co1212\cite{Huang}. Ga1212\cite{Roth} and
Al-1212\cite{Ramirez,Bordet} have also been reported to have the
tetrahedral polyhedron. Such $CoO_4$ tetrahedra forms a zigzag
chain running diagonally relative to the pervoskite base. A
regular alternation of two zigzag chains, which are mirror images
of each other, forms an orthorhombic superstructure\cite{Nagi}.
Unlike $CuO_4$ square chains in Cu1212, which is very easy to lose
or obtain oxygen, the $CoO_4$ tetrahedral chains in Co1212 are
reported to be a rigid configuration in terms of oxygen
content\cite{Morita2}. In Co1212, all of the Co ions reside on the
Cu(1) sites\cite{Huang}. With $Y^{3+}$ substituted by $Ca^{2+}$,
however, Co and Cu will intermixed partly. The M(T) curve of
Co1212 exhibits a broad maximum, indicative of a low-dimensional
magnetic ordering, which was thought to originating from $Co^{3+}$
spin\cite{Awana}. With no doubt, studies on the function of
$CoO_4$ tetrahedra certainly give very useful information to
settle the forenamed question. To our knowledge, no report has
been reported on the transport properties of
$CoSr_2Y_{1-x}Ca_xCu_2O_{7+\delta}$. In the present report, we
investigated the magnetic and electrical properties on Ca
substituted Co1212 from Ca content 0.0 to 0.4. It is found that
with increasing Ca doping level, electrical transport properties
become related to magnetic structure closely. Magnetic properties
and their correlation with charge transport suggest that slightly
partial holes induced by Ca doping reside on the $CoO_{1+\delta}$
layers and these holes influence the magnetic and charge transport
properties dramatically. In addition, these holes cause
$CoO_{1+\delta}$ to participate in bulk electrical transport.
X-ray photoelectron spectroscopy(XPS) studies indicate that for
high Ca doping level the valence of Co increases comparing to the
low doping level and thus confirm that holes enter into the charge
reservoir.

\section{EXPERIMENT}
A series of samples with compound of
$CoSr_2Y_{1-x}Ca_xCu_2O_{7+\delta}$($x$=0.0$\sim$0.4) were
synthesized through a solid-state reaction route. $Co_2O_3$,
$SrCO_3$, $Y_2O_3$, $CaCO_3$, and $CuO$, as the starting
materials, were mixed in nominal composition and heated for 24h at
950 $^o$C and another 24 h at 1000 $^o$C with intermediate
grinding. Finally, the products were pelletized and sintered at
1010 $^o$C for 24h. The phase purity was checked for each sample
with powder X-ray diffraction(XRD). Magnetization measurements
were performed on each sample with a SQUID magnetometer in field
cooling mode. For the sample $x$=0.25 and $x$=0.4, additional ZFC
(zero field cooling) magnetization were performed. Resistance
measurements were performed using AC four-probe method with AC
bridge resistance bridge system(Linear Research Inc.; LR-700P).
All samples for magnetization and resistance measurements have
been annealed under 195 atm oxygen atmosphere at 500 $^o$C for
24h. It should be pointed out that the sample annealed under 195
atm and ambient pressure oxygen atmosphere shows almost the same
behavior in charge transport and magnetism. These results further
confirm that the oxygen content in Co1212 cannot change easily.
The XPS spectra were collected with an ESCALAB MK II electron
spectrometer using Mg K$\alpha$ radiation as excitation source
($h\nu$=1253.6 eV). Binding energy was calibrated with a reference
of surface contaminated C 1s (E$_b$=284.6 eV). The basic vacuum
was about 1$\times$10$^{-9}$ mbar.

\section{RESULTS and DISCUSSION}
XRD patterns for as-synthesized(AS) samples with various Ca
content, $x$, are shown in Fig. 1.   No any visible impurity peak
in XRD patterns is observed for all samples. All the peaks can be
indexed using the space group $Ima2$ \cite{Huang}. It demonstrates
all the samples are single phase materials and have the
orthorhombic structure. The lattice parameters determined from the
XRD patterns with the Rietveld method in GSAS program are plotted
against $x$ in Fig. 2 (Noted that the c-axis here were referred to
the layer piling direction as typically employed for the
structures of high-T$_c$ superconductors, and a-axis is along the
$CoO_{1+\delta}$ chain.). With increasing Ca content $x$, the
lattice parameter of a-axis slightly increases, while b-axis
shrinks a little. An increase of c-axis parameter with x
demonstrates that the doped Ca ions have entered into Y sites
instead of Sr sites. The reason is that the radius of $Ca^{2+}$ is
larger than $Y^{3+}$ but less than $Sr^{2+}$, thus substitution of
$Sr^{2+}$ by $Ca^{2+}$ should cause a shrinkage of c-axis. It has
been reported that the oxygen content in Ca-doped Co1212 almost
remains constant\cite{Morita}, thus replacing partial trivalent Y
ion by divalent Ca ion results in excess holes to system.

Figure 3 shows the temperature dependence of the dc magnetization
(M) of annealed samples $x$=0.0$\sim$0.4. The magnetization of
Ca-free sample shows a broad maximum around 145 K and then a
shallow down-turn followed by a up-turn with decreasing
temperature, which well agrees with the previous
reports\cite{Awana,Awana1}. The analogous FC M(T) curve has been
observed in $Sr_2CoCuS_2O_2$\cite{Matoba},
$Sr_2CuMnO_3S$\cite{Kamihara}, $Sr_2MnO_3Cl$ and
$Sr_4Mn_3O_{8-y}Cl_2$\cite{Knee}, in which the broad maximum was
thought to be indicative of low-dimensional magnetic
feature\cite{Navarro}. The inverse of magnetization displays an
upturn deviation from linear behavior with decreasing temperature,
indicative of antiferromagnetic ordering. Considering the 3D
magnetic character of high-T$_c$ system, the low dimensional
antiferromagnetic feature most likely originates from the
 Co-O layers instead of Cu-O planes. In addition, magnetization
 remains almost unchange
for $x$=0.1 and $x$=0.2 except for a little lowering of the
temperature of maximum, indicating that the holes induced by Ca
doping enter into the Cu-O plane. The drastic drop of resistivity
with $x$ increasing from 0.0 to 0.2 confirms this speculation,
which will be discussed later. As shown in fig.3, for higher Ca
doping, however, the temperature of maximum decreases
dramatically, and for the sample with $x$=0.4, there only exhibits
a small kink in M(T) curve. From Fig. 3b, the inverse of M(T) of
sample $x$=0.4 shows a downturn deviation from high temperature
linear behavior with decreasing temperature, which indicates that
at low temperature the sample $x$=0.4 shows a weak ferromagnetism.
The branching of the ZFC and FC data shown in the inset of figure
3 approves the ferromagnetism in the sample $x$=0.4. The change of
magnetic behavior by Ca-doping may arise from following aspects:
1) some holes enter into $CoO_{1+\delta}$ and destruct the
effective antiferromagnetic coupling; 2) intermixing of Cu and Co
ion induced by Ca doping \cite{Huang} breaks the Co-O chain; 3)
some structure transformation changes the Co coordination. The
following results will suggest that the first supposition may be
the most possible.

Resistivity as a function of temperature is displayed in Fig. 4
for the annealed samples $x$=0.0$\sim$0.4. Substitution of Y by Ca
reduces the resistivity fiercely with magnitude change of three
orders at room temperature from $x$=0.0 to 0.4. A  metal-insulator
transition induced by doping Ca is clearly observed with
increasing doping level. In contrast to the large resistivity and
insulator-like behavior in Ca-free Co1212 sample, the resistivity
of the sample $x$=0.2 is reduced dramatically by nearly 1000 times
at room temperature, and it exhibits a metallic behavior with a
positive slope within a wide temperature region below room
temperature. Doping Ca up to $x$=0.4, the resistivity at room
temperature is very close to that of the superconducting cuprates.
Since the conductive unit in high-T$_c$ cuprates is $CuO_2$ plane,
drastic drop of resistivity induced by Ca substitution for Y ion
demonstrates that doped holes enter into $CuO_2$ plane indeed. The
variation of resistivity with doping is  associated with the
magnetic results discussed above. However, none of all the samples
shows superconducting (SC) at the temperature down to 2 K. The
reason for absence of SC has been ascribed to that slight Co ions
enter into $CuO_2$ plane and this destroys its integrality
\cite{Huang} or that some holes are trapped in the
$CoO_{1+\delta}$ charge reservior \cite{Morita}. It is excluded
for the former interpretation since it can be become
superconducting by annealing AS samples in 5 GPa oxygen
\cite{Morita} (this seems to be unable to rearrange the
occupational sites). Otherwise,  our results seem to demonstrate
that the charge reservoir participates in charge transport, that
is, there exist holes in $CoO_{1+\delta}$ layers.

In Fig. 4, there exists a small peak at low temperature for the
samples of $x$=0.2, 0.25, 0.3 and 0.4  with a noticeable change in
the slope ($d\rho(T)/dT$). In Fig. 5, the derivative of the
resistivity for the four samples $x$=0.2, 0.25, 0.3, 0.4 shows a
shallow dip-hump feature. It is found that the temperature at
maximum of M(T) corresponds to this dip-hump zone for these four
samples. It suggests that the peak observed in resistivity is
related to the maximum of the magnetization M(T). In order to find
out the correlation between resistivity and magnetization,  the
curvature ($d^2\rho(T)/dT^2$) and M(T) curves are shown in Fig.6.
The temperatures of maximum of M(T) are almost the same as that
corresponding to the minimum of $d^2\rho(T)/dT^2$ for the four
samples, with the largest difference less than 10 K for the sample
$x$=0.2. These results indicate that when Ca content $x$ is as
large as 0.2, electrical transport behavior becomes related to
magnetic properties closely.

What is the nature of this correlation between charge transport
and magnetic behavior? As mentioned previously, the $CuO_4$ chains
participate in the b-axis charge transport in Cu1212, and give
slightly contribution to the resistivity and lead to a different
temperature dependence from that of the a-axis. In our case,
magnetic properties in this system originate from the
$CoO_{1+\delta}$ charge reservoir, and the main conducting unit is
$CuO_2$ plane. Two possible reasons can be considered to interpret
the fact that the charge transport is closely related to magnetic
properties: $i$)the couplings between the itinerant charge carrier
in the $CuO_2$ planes and the localized ordered spin in the
$CoO_{1+\delta}$ charge reservoir lead to the correlation;
$ii$)likely to Cu1212, the charge reservoir participates in charge
transport, so that magnetic transition occurred in
$CoO_{1+\delta}$ influences the charge transport. The behavior
exhibited in Co1212 is not the former case because the correlation
between magnetic and charge transports is different from that in
Ru1212 and Ru1222, in which there exists a ferromagnetic
transition in $RuO_2$ plane, but no apparent change in resistivity
is observed at Curie temperature \cite{McCrone, Chen}. This
assumption is further tested by the magnetotransport data
discussed later. In this way, the system contains contributions
from two types of conducting layer: $CoO_{1+\delta}$ layers and
$CuO_2$ planes although the component arising from the charge
reservoir layers may be rather little. Therefore, an apparent
change in resistivity is observed when a magnetic
 transition takes place in $CoO_{1+\delta}$, then the dip-hump structure on
d$\rho{T}$/dT is related to magnetic transition on the
charge reservoir layers.

This speculation is confirmed by the magnetotransport data. As
shown in Fig. 7, the sample $x$=0.3 exhibits positive
magnetoresistance (MR), less than 1$\%$ at magnetic field as high
as 14 Tesla, while the sample $x$=0.4 shows a negative MR, larger
than 14$\%$ at 14 Tesla, which seems to be like the behavior in
ferromagnetic metal. These contrasting behaviors manifest that the
magnetism is transformed from low-dimensional antiforremagnetism
in the sample $x$=0.3 to weak ferromagnetism in the sample
$x$=0.4, a different MR behavior occurs and the change in
resistivity behavior with Ca doping is closely associated with
that in magnetism. Furthermore, the different MR behavior
associated with the different magnetic properties located on the
$CoO_{1+\delta}$ layers further confirms that the $CoO_{1+\delta}$
charge reservoir participates in charge transport. In Ru-1212 and
Ru1222, a maximum negative MR is observed at $T_{curie}$
(Ferromagnetic transition temperature) \cite{McCrone, Chen}.
 This MR behavior has been attributed to the interaction between Ru moment in $RuO_2$
planes and itinerant carriers in $CuO_2$ planes \cite{McCrone,
Chen}. However, the sample $x$=0.4 in Co1212 shows a different MR
behavior from Ru1212 and Ru1222. An increase in negative MR with
decreasing temperature is observed in Co1212 and the MR in Co1212
is much larger than that in Ru1212 and Ru1222. These differences
between Co1212 and Ru1212 (Ru1222) is also an indication for
significant current flowing in $CoO_{1+\delta}$ layers.
 The Kohler's plot for the samples is shown in the inset of figure 7. It should be noted that
the MR of the samples with $x$=0.3 and 0.4 does not obey Kohler's
rule. This is because below 110 K the charge becomes localized
instead of a metal as above 110 K. The MR of the sample $x$ =0.4
is proportional to square of magnetic field, while that of the
sample $x$=0.3 shows saturation at high field. In addition, the
evolution of the magnetic properties with Ca doping supports minor
charge carrier in $CoO_{1+\delta}$ layers.
 For the samples  with  Ca content $x\leq 0.2$, the holes induced by Ca substituting for Y are
completely introduced into $CuO_2$ planes, so that the resistivity
decreases apparently and the magnetic behavior originated from
$CoO_{1+\delta}$ layers does not change with Ca doping. However,
further doping Ca leads to a dramatic change in the magnetic
behavior and an associated change in resistivity with magnetic
behavior. It suggests that the change of magnetic properties is
induced by the transfer of the holes into the $CoO_{1+\delta}$
layers when the Ca content $x > 0.2$. It should be pointed out
that the ferromagnetism occurred in $RuO_2$ layers does not change
with the doping level.

The consideration that holes enter into the $CoO_{1+\delta}$
charge reservoir for the high Ca doping samples is confirmed by
XPS results further. The Co2p spectra obtained for the sample
$x$=0.1 and $x$=0.4 were shown in figure 8. Comparing to the
spectrum of the sample $x$=0.1, the binding energy of the two main
components Co $2p_{3/2}$ and $2p_{1/2}$ in the spectrum for the
sample $x$=0.4 shift toward higher energy as shown by the solid
straight line in the figure. The spectra were analyzed using a
peak synthesis program in which a non-linear background is assumed
\cite{Shirley}. For the sample $x$=0.1, Co2p spectrum shows two
main components at  about 779.3 eV and 794.8 eV which are
attributed to Co$^{3+}$ ions \cite{Dupin}. In comparison with the
spectrum of the sample $x$=0.1, The two peaks corresponding to the
two components Co $2p_{3/2}$ and $2p_{1/2}$ become broad,
especially, an apparent doublet is observed for the component Co
$2p_{1/2}$. The peaks in the spectrum of $x$=0.4 have to be fitted
by two components. By fitting, the doublet positions are
determined to be at 797.0 and 795.3 eV for the Co $2p_{1/2}$, and
at 781.2 and 779.8 eV for the Co $2p_{3/2}$, respectively. As
pointed out in Ref. 24, the binding energies of 779.8 and 795.3 eV
correspond to Co$^{3+}$ ions, while 781.2 and 797.0 eV are
attributed to Co$^{4+}$ ions. This is evident by the presence of
an obvious shoulder on the high-energy side of the Co $2p_{1/2}$
component (marked by the arrow in the figure). These facts clearly
show the increase of the valence of Co and confirm the transfer of
the holes into the $CoO_{1+\delta}$ layers by increasing the Ca
doping level.

It has been reported that $CoO_4$ tetrahedra in Ca-free Co1212 has
a rigid configuration in terms of the oxygen content, and
annealing in flowing Ar or $O_2$ have little effect on oxygen
stoichiometry \cite{Morita2}. It is found that no obvious
difference in resistivity value or its behavior is observed in AS,
175 atm oxygen annealing, and 195 atm oxygen annealing samples for
$x$=0.1, 0.2, 0.25, 0.3 and 0.4 (not shown here). From the
analysis on magnetic and transport behavior above, doping of Ca
does induce holes into $CoO_{1+\delta}$ although the amount of
these holes may be much smaller than that of holes entering into
$CuO_2$ planes for the same Ca doping level. From iodometric
titration experiment by Morita et al. \cite{Morita}, the oxygen
content almost remains constant, thus holes entering into
$CoO_{1+\delta}$ increase the oxidation state of Co. When magnetic
ordering occurs in insulating $Co^{3+}$-O matrix, the magnetism
occurred in $CoO_{1+\delta}$ layers has very weak effect on the
charge transport. The magnetic effect on resistivity originates
from the coupling between the itinerant carrier and the local
ordered spin in the $CoO_{1+\delta}$ layers. When hole is doped
into $CoO_{1+\delta}$ layers, ferromagnetic metal (FM) cluster
containing $Co^{4+}$ ions may reduce the resistivity apparently,
especially for the sample $x$=0.4 in which weak ferromagnetism
shows up. This picture has been used to interpret the noticeable
change of the slope in $\rho(T)$ of $La_{0.85}Sr_{0.15}CoO_3$ and
$La_{0.75}Sr_{0.25}CoO_3$ films \cite{Prokhorov}. In fact, in the
inset of figure 3, the ZFC and FC magnetization curves of the
samples $x$=0.25 and 0.4 start branching around the temperatures
corresponding to the maximum of magnetization, indicative of some
ferromagnetic component. However, as reported in ref.11, for the
Ca-free sample no ZFC and FC branching are observed. These results
suggest that Ca-free sample contains no ferromagnetic component,
while as Ca doping level is higher than 0.2, the system becomes a
intermixture of antiferromagnetic and ferromagnetic component. The
temperature dependence of the magnetization for the sample
$x$=0.2, 0.25, 0.3 exhibit antiferromagnetism, indicating that in
these samples ferromagnetic component is rather small. While for
the sample $x$=0.4, it exhibits weak ferromagnetism, indicating
that in this sample the ferromagnetic component is already
preponderant. This may interpret the large MR in the sample
$x$=0.4 while rather small MR in the sample $x$=0.3. It is not
clear whether the partly intermixing between Co and Cu has any
effect on the resistivity. More microscopic studies on the
electronic state of Co ions($Co^{3+}$ or $Co^{4+}$) in the charge
reservoir is needed to explain this phenomena completely.

\section{conclusion}
We have synthesized polycrystalline
$CoSr_2Y_{1-x}Ca_xCu_2O_{7+\delta}$($x$=0.0$\sim$0.4), and
investigated their magnetic and transport properties. In the
sample $x$=0.0$\sim$0.2, magnetization shows broad maximum at low
temperature (almost the same temperature), indicative of
low-dimensional antiferromagnetic nature originated from
$CoO_{1+\delta}$ layers. The temperature of such maximum for the
sample $x$=0.3 drops dramatically, and for the sample with $x$=0.4
a weak ferromagnetism  shows up. Ca doping reduces resistivity
quickly. d$\rho$(T)/dT shows dip-hump feature at low temperature
for the samples with $x$=0.2$\sim$0.4, which is ascribed to a
contribution of holes in $CoO_{1+\delta}$ charge reservoir and
magnetic transition within the layers. The speculation is
confirmed by the fact that small positive MR exhibits in the
sample $x$=0.3,while the large negative MR in the $x$=0.4 sample.

\section{ACKNOWLEDGEMENTS}
 This work is supported by the grant from the Nature
Science Foundation of China and by the Ministry of Science and
Technology
of China (Grant No. NKBRSF-G1999064601), the Knowledge Innovation Project of Chinese Academy of Sciences.\\

\newpage

\noindent
{\bf FIGURE CAPTIONS} \\

\noindent
 Figure 1 :
 XRD patterns for the as-synthesized samples of
$CoSr_2Y_{1-x}Ca_xCu_2O_{7+\delta} $($x$=0.0$\sim$0.4).\\

Figure 2:
Lattice parameters for the as-synthesized
samples of $CoSr_2Y_{1-x}Ca_xCu_2O_{7+\delta}$.\\

Figure 3: (a) The temperature dependence of magnetization measured
for the annealed $CoSr_2Y_{1-x}Ca_xCu_2O_{7+\delta}$ samples in
field-cooled process at the field of 100 Oe. Inset: ZFC (dashed
line) and FC (solid line) data at the field of 100 Oe plotted
against the temperature for the samples $x$=0.25 and 0.4.
(b) The temperature dependence of inverse magnetization for the sample $x$=0.0$\sim$0.4.\\

Figure 4: The resistivity as a function of temperature for the
annealed
$CoSr_2Y_{1-x}Ca_xCu_2O_{7+\delta}$ samples.\\

Figure 5:
The resistivity and its derivative as a function of temperature for the samples with $x$=0.2, 0.25, 0.3, 0.4.\\

Figure 6: The temperature dependence of the resistivity
curvature(d$^2\rho$/d$^2$T)
and magnetization for the samples with $x$=0.2, 0.25, 0.3, 0.4.\\

Figure 7: The plot of magnetoresistance
(MR=[$\rho$(T,H)-$\rho$(T,0)]/$\rho$(T,0)$\times$100$\%$ ) vs.
magnetic field($\mu_0$H) for the samples with  $x$=0.3, 0.4.
Insets: the same data are shown in Kohler's plot.\\

Figure 8: The Co2p XPS spectra for the sample
$x$=0.1 and 0.4.\\

\end{document}